\documentclass[journal]{IEEEtran} 

\usepackage{fb,afterpage} 
\labelfigure{./}
\graphicspath{{./}}

\usepackage{tikz}
\usetikzlibrary{arrows}
\usepackage[linesnumbered,lined, algoruled]{algorithm2e}

\definecolor{tomblue}{RGB}{0,0,255}
\def\Tom#1{{#1}}
\def\snr{\Tom{\Omega}}
\begin{document}

\title{On the Evaluation of the Polyanskiy-Poor-Verd\'u Converse Bound for Finite Block-length Coding in AWGN}

\author{%
Tomaso~Erseghe
\thanks{Author is with Dipartimento di Ingegneria dell'Informazione, Universit\`a di Padova, Via G. Gradenigo 6/B, 35131 Padova, Italy. 
} %
\thanks{\copyright 2015 IEEE. Personal use of this material is permitted. Permission from IEEE must be obtained for all other uses, including reprinting/republishing this material for advertising or promotional purposes, collecting new collected works for resale or redistribution to servers or lists, or reuse of any copyrighted component of this work in other works.}
} 

\maketitle


\acrodef{AWGN}{additive white Gaussian noise} 
\acrodef{BSC}{binary symmetric channel} 
\acrodef{CDF}{cumulative distribution function} 
\acrodef{CCDF}{complementary cumulative distribution function} 
\acrodef{FA}{false alarm} 
\acrodef{LDPC}{low density parity check} 
\acrodef{MD}{missed detection} 
\acrodef{ML}{maximum likelihood} 
\acrodef{PAM}{pulse amplitude modulation} 
\acrodef{PDF}{probability density function} 
\acrodef{SNR}{signal to noise ratio} 
\acrodef{PPV}{Polyanskiy, Poor, and Verd\'u}

\begin{abstract}
A tight converse bound to channel coding rate in the finite block-length regime and under AWGN conditions was recently proposed by \ac{PPV}. The bound is a generalization of a number of other classical results, and it was also claimed to be equivalent to Shannon's 1959 cone packing bound. Unfortunately, its numerical evaluation is troublesome even for not too large values of the block-length $n$. In this paper we tackle the numerical evaluation by compactly expressing the \ac{PPV} converse bound in terms of non-central chi-squared distributions, and by evaluating those through a \Tom{an integral expression and a corresponding} series expansion \Tom{which exploit a method} proposed by Temme. As a result, a robust evaluation method and new insights on the bound's asymptotics, as well as new approximate expressions, are given.
\end{abstract}

\begin{IEEEkeywords}
Channel capacity, Coding for noisy channels, Converse, Finite block-length regime, Shannon theory.
\end{IEEEkeywords}

\section{Introduction}

Recently, a number of new results for channel coding performance in the (non asymptotic) finite block-length regime have been proposed by Polyanskiy, Poor, and Verd\'u \cite{Polyanskiy10,polyanskiy2010channel}. Of particular interest in their work is a converse theorem \cite[Theo.~27]{Polyanskiy10}, which was proved to generalize classical results. Its application to the \ac{AWGN} channel was shown to be as accurate as the 1959 Shannon's bound based on cone packing \cite[\Tom{error probability lower bound in equation (3)}]{Shannon59}, while a formal claim of the identity between the two approaches is available in \cite{Polyanskiy13}. Numerical evaluation of \Tom{both Shannon's and \ac{PPV} bounds is challenging. The numerical difficulties involved in calculating Shannon's result are discussed in \cite{wiechman:it08}. The \ac{PPV} bound instead} relies on $\chi^2$ (chi-squared) distributions which are recognized in \cite{Polyanskiy10} to be hard to evaluate even for not too large values of $n$. As a matter of fact, a number of bounding techniques (not explicitly described in the paper) have been used for numerical evaluation, and, concurrently, a simple asymptotic expression has been identified by using a normal approximation approach.

\Tom{It should be clarified that the above is only a partial view of the many approaches that have been used over the years to identify the performance limits in the finite block-length regime. To cite a few alternative methods, we recall the classical error exponent approach \cite{gallager1968information}, channel dispersion type of approximations \cite{tan2015it}, infinite constellation results \cite{Poltyrev:it94,Ingber:it13}, and also moderate deviations \cite{altug2010moderate}. }

In this paper we wish to provide new insights on the \ac{PPV} converse bound by overcoming the difficulty of numerically evaluating  it in \ac{AWGN} conditions. We therefore propose to exploit the results of Temme \cite{Temme93}, providing a series expansion for $\chi^2$ distributions which is meaningful for large values of $n$. \Tom{A single-integral expression is also made available for numerical calculation, together with a method to evaluate the accuracy of the proposed asymptotic expansion.} In this way we are able to identify reliable methods for evaluating the bound, and, furthermore, \Tom{novel} simple expressions to fully capture its asymptotic behavior. 

The paper is organized as follows. In \sect{SC} we depict the scenario of interest, assess notation, and provide the general formulation for the converse bound, as well as its explicit formulation in terms of $\chi^2$ distributions. Then, \sect{AW} exploits \cite{Temme93}, and presents an efficient method for evaluating the \ac{PPV} bound. Application examples with meaningful performance measures and asymptotic behavior are discussed in \sect{EX}. To keep the flow of discussion, all theorems proofs are collected in the Appendix.

\Section[SC]{The Converse Bound}

\subsection{Notation for the \ac{AWGN} scenario}

We assume a standard communication scenario in \ac{AWGN}. The transmitted message $W\in\{1,\ldots,M\}$ is first encoded into a real-valued vector $\B x = \B c_W$ of length $n$. \Tom{The codewords set (or simply the code) is denoted with $\C C = \{\B c_1,\ldots,\B c_M\}$, and} the information rate with $R=\frac1n\log_2 M$. Codewords are assumed to belong to set \Tom{
$$
\C F=\Big\{ \B x\Big| \|\B x\|^2=n\sigma_x^2\Big\}\subset \M R^n\;,
\e{MO28}
$$
namely a constant energy set where $\sigma_x^2$ is the average transmitted power, i.e., the power per symbol. Equation \e{MO28} sets an equal-power constraint. Generalization of the result to a maximum or average power constraint can be obtained by exploiting \cite[Lemma~65, \S 4.2]{polyanskiy2010channel}, but see also \cite[Theorem~77, \S 4.3.3]{polyanskiy2010channel}.} 

The transmission channel maps the transmitted codeword into a real-valued received vector $\B y$ of length $n$. The (memoryless) channel is fully described by transition probabilities 
$$
p_{y|x}(\B b|\B a) = \prod_{i=1}^n \frac1{\sqrt{2\pi \sigma_w^2}} e^{-\frac12\frac{(b_i-a_i)^2}{\sigma_w^2}}\;,
\e{MO4}
$$
with $\sigma_w^2$ the noise variance. At the receiver side, a decoding algorithm is applied to extract from $\B y$ an estimate $\widehat{W}$ of the transmitted message. The error probability is denoted with $P_e ={\rm P}[\widehat{W}\neq W] $.

For a given choice of the \ac{PDF} of transmitted codewords $p_x(\B a)$, we denote the joint \ac{PDF} with $p_{xy}(\B a,\B b)=p_{y|x}(\B b|\B a)p_x(\B a)$, and the received vector \ac{PDF} with $p_y(\B b) = \int p_{y|x}(\B b|\B a)p_x(\B a)d\B a$. 
\Tom{The capacity} $C$ is achieved for a Gaussian input distribution with associated symbol variance $\sigma_x^2$, to have
$$
C = \fract12\log_2(1+\snr)\;,\qquad \snr = \frac{\sigma_x^2}{\sigma_w^2}\;,
\e{MO5}
$$
where $\snr$ is the reference \ac{SNR} at the receiver side. It also is $\sigma_y^2 = \sigma_x^2+\sigma_w^2 = (1+\snr)\sigma_w^2$, with $\sigma_y^2$ the received signal variance. 

For later use we also introduce the  Gaussian \ac{CCDF} $Q(x)$. For $x>0$ we write it in the form $Q(x) = e^{-\frac12x^2}q(x)$ where (see, e.g., \cite[(7.1.23)]{AbramowitzStegun68})
$$
q(x) = \frac1{2\sqrt{\pi}}\sum_{k=0}^\infty (-1)^k \frac{\Gamma(k+\frac12)}{\Gamma(\frac12)} \left(\frac{2}{x^2}\right)^{k+\frac12} \;,
\e{TE10c}
$$
\Tom{and where $\Gamma(\cdot)$ is the gamma function. For $x<0$ we have} $Q(x)=1+e^{-\frac12x^2}q(x)$.

\subsection{Notation for binary hypothesis testing}

Bounds are related to the performance of a Neyman-Pearson test, namely the uniformly most powerful test in discriminating between two hypotheses (see, e.g., \cite{kay:93}). The test of interest is between hypotheses
$$
\eqalign{
\C H_1\;:&\qquad \B y\sim p_{y|x}\cr
\C H_0\;:&\qquad \B y\sim q_y\;,
}
\e{MO8}
$$
for a given $\B x$, and for some choice of the \ac{PDF} $q_y$. In this context the Neyman-Pearson test is the threshold test that discriminates between $\C H_0$ and $\C H_1$ by inspecting the log-likelihood function
$$
\Lambda(\B x,\B y) = \frac1n\ln \frac{p_{y|x}(\B y|\B x)}{q_y(\B y)}\;.
\e{MO10}
$$
Hypothesis $\C H_1$ is selected if $\Lambda(\B x,\B y)\ge\lambda$, with $\lambda$ a given threshold, and hypothesis $\C H_0$ is selected otherwise. Probabilities of interest are \ac{MD} and \ac{FA} probabilities which will be denoted, respectively, with 
$$
\eqalign{
P\sub{MD}(\B x,\lambda) & = \P{\Lambda(\B x,\B y)<\lambda\Big|\C H_1,\B x}\cr
P\sub{FA}(\B x,\lambda) & = \P{\Lambda(\B x,\B y)\ge\lambda\Big|\C H_0,\B x}\;,
}
\e{MO12}
$$
to underline their dependency on the values of both $\B x$ and $\lambda$. 

The Neyman-Pearson test applied to hypotheses
$$
\eqalign{
\C H_1\;:&\qquad (\B x,\B y)\sim p_{y|x}p_x\cr
\C H_0\;:&\qquad (\B x,\B y)\sim q_yp_x\;,
}
\e{MO18}
$$
is also built on log-likelihood function \e{MO10}, and its \ac{MD} and \ac{FA} probabilities satisfy the average relations
$$
\eqalign{
P\sub{MD}(\lambda) & =\sum_{\Bs a\Tom{\in\C C}}  P\sub{MD}(\B a,\lambda) \,p_x(\B a)\cr
P\sub{FA}(\lambda) & =\sum_{\Bs a\Tom{\in\C C}} P\sub{FA}(\B a,\lambda)\,p_x(\B a)\;.
}
\e{MO20}
$$

\Tom{Note that \e{MO20} is in general dependent on the chosen code $\C C$. However, a result independent of $\C C$ is obtained under the assumption that both $P\sub{MD}(\B x,\lambda)$ and $P\sub{FA}(\B x,\lambda)$ are independent of $\B x$, for $\B x\in\C F$, so that \e{MO12} and \e{MO20} coincide. This is the case for the \ac{AWGN} channel} when the auxiliary statistical description $q_y$ is set to the capacity achieving output distribution. The result is provided by the following theorem, which is a reformulation of \cite[Theorem~40]{Polyanskiy10} to explicit the role of non-central $\chi^2$ distributions. 

\begin{theorem}\label{teoPOLawgn}
An \ac{AWGN} channel with constant energy codewords \Tom{$\C C\subset\C F$}, and with $q_y\sim\C N(\B 0,\B I_n\sigma_y^2)$, has Neyman-Pearson \ac{FA} and \ac{MD} probabilities \Tom{\e{MO12} which are independent of $\B x\in\C F$, and which are given by}
$$
\eqalign{
P\sub{MD}(\lambda) & = P\sub{MD}(\B x,\lambda)  = \overline{F}_\chi\left(n\lambda';n,\frac n\snr\right)\cr
P\sub{FA}(\lambda)  & = P\sub{FA}(\B x,\lambda)= F_\chi\left(\frac{n\lambda'}{1+\snr};n,n\frac{1+\snr}{\snr}\right)\;,
}
\e{MO30}
$$
where $F_\chi(a;n,s)$ and $\overline{F}_\chi(a;n,s)$ denote, respectively, the $\chi^2$ \ac{CDF} and \ac{CCDF} of order $n$ and parameter $s$. In \e{MO30}, $\lambda$ and $\lambda'$ are in the linear relation
$$
2\lambda = 1+\ln(1+\snr)-\frac{\snr}{(1+\snr)}\lambda'\;.
\e{LL2}
$$
\hfill~$\Box$
\end{theorem}

\subsection{Converse theorem in \ac{AWGN}}

The converse bound is given in \cite[Theorem~28]{Polyanskiy10} \Tom{(average probability of error) and  \cite[Theorem~31]{Polyanskiy10} (maximal probability of error)}. Without any loss in generality, we reshape it into a more explicit form making direct use of Neyman-Pearson \ac{FA} and \ac{MD} probabilities. For completeness, a (very short) proof is available in the Appendix.

\begin{theorem}\label{achieveandconverse}
Assume that: a) the block-length $n$ is finite; b) the input messages are equally likely, \Tom{$p_x(\B a)=\frac1M$, $\B a\in\C C$}; c) the error probability is $P_e$ \Tom{in either the average or the maximum probability sense}; d) \Tom{$q_y$ is chosen in such a way that Neyman-Pearson \ac{FA} and \ac{MD} probabilities \e{MO12} are independent of $\B x\in\C F$}. Then the code rate $R$ satisfies
$$
R \le \overline{R}  = -\frac1n\log_2 P\sub{FA}(\lambda)\;,
\e{MO26}
$$
where $\lambda$ is set by constraint $P\sub{MD}(\lambda)=P_e$.\hfill~$\Box$
\end{theorem}
\Tom{In the \ac{AWGN} scenario the natural choice for $q_y$ is the Gaussian choice of Theorem~\ref{teoPOLawgn}, in which case the \ac{FA} and \ac{MD} probabilities are given by \e{MO30}.}

Observe that, as a direct consequence of the definition of \ac{FA} and \ac{MD} probabilities, the bound $\overline{R}$ is ensured to be non-decreasing in $\epsilon$, and, similarly, the value of $\lambda$ is also non-decreasing in $\epsilon$. Observe also that the result can be used to obtain a bound to the error probability $P_e$ for a given rate $R$ and block-length $n$. Specifically, it is
$$
P_e \ge\underline{P}_e =  P\sub{MD}(\lambda)
\e{MO27}
$$
where $\lambda$ is set by constraint $-\frac1n\log_2P\sub{FA}(\lambda)=R$. This bound is useful for practically assessing the performance of specific encoder/decoder choices.

\Section[AW]{Numerical Evaluation of \Tom{\ac{FA} and \ac{MD} Probabilities}}

The numerical evaluation of $\chi^2$ \acp{CDF} and \acp{CCDF} in \e{MO30} is troublesome, especially for large values of the order $n$. To overcome this difficulty we exploit the results of \cite[(4.2)-(4.3)]{Temme93} and obtain \Tom{an integral form and} a series expansion which are meaningful (and applicable) in the present context. The series is in close relation with standard asymptotic expansions for the Gaussian \ac{CCDF} \e{TE10c}. \Tom{Incidentally, note that other existing results which could be useful in principle in our context, e.g., the bounds on the Marcum $Q$-function available in \cite{corazza2002new}, are too loose to provide significant results.}

\subsection{\Tom{Integral form}}

\Tom{The results of \cite[(4.2)]{Temme93} provide the following formalization for \ac{FA} and \ac{MD} probabilities.}

\begin{theorem}\label{teoTE0}
\ac{FA} and \ac{MD} probabilities \e{MO30} can be expressed in the form
$$
\eqalign{
P\sub{MD} &  =  1(s\sub{MD}(\gamma)) +  g\sub{MD}(\gamma)\,e^{- \frac12 n v\sub{MD}(\gamma)} \cr
P\sub{FA} &  =  1(-s\sub{FA}(\gamma)) -  g\sub{FA}(\gamma)\,e^{- \frac12 n v\sub{FA}(\gamma)}\;,
}
\e{TE10}
$$
where $1(\cdot)$ is the unit step function providing $1$ for $x\ge0$ and $0$ otherwise, and threshold $\gamma>0$ is related to $\lambda'$ by 
$$
\lambda'  = \snr/(4s^2_\gamma) \;,\quad
\Tom{s_\gamma = \sinh(\gamma)}\;. 
\e{TE10a}
$$
\Tom{Depending on wether we are dealing with the \ac{MD} or the \ac{FA} case, and with a little abuse of notation}, functions $s$, $g$ and $v$ in \e{TE10} are defined by 
$$
\eqalign{
\theta(\gamma) & =  \ln(\snr/(2s_\gamma)) +  \cases{ 0 & MD \cr -\ln(1+\snr)& FA} \cr
s(\gamma) & ={\rm sign}(\gamma-\theta(\gamma)) \cr 
v(\gamma) & = -\alpha(\theta(\gamma))/s_\gamma\cr
\alpha(x) &= c_\gamma-\cosh(x)+s_\gamma\,(x-\gamma)\;,\quad \Tom{c_\gamma=\cosh(\gamma)}\cr
g(\gamma) & = \Tom{\frac1\pi \int_{0}^\infty \Im[f(u)]\, e^{-nu^2/(4s_\gamma)} \;du}\cr
f(u) & = \frac{t'(u)}{e^\theta-t(u)}
}
\e{TE4_v0}
$$
where $t(\cdot)$ is defined by the inverse relation
$$
\Tom{t^{-1}(x)  = {\rm sign}(\Im(x))\,\sqrt{2\alpha(\ln(x))}}\;,
\e{NM4}
$$
\Tom{and where $\Im$ extracts the imaginary part}.\hfill~$\Box$
\end{theorem}

\Tom{The definition of function $g(\gamma)$ in \e{TE4_v0} is targeted to an asymptotic expansion of the result, but can be suitably mapped into a form which is more convenient for numerical evaluation of the integral. We therefore exploit the steepest descent path defined in \cite{Temme93} to obtain the following result.}

\Tom{\begin{theorem}\label{teoTE0b}
Function $g(\gamma)$ in \e{TE4_v0} can be equivalently expressed in the form
$$
g(\gamma) = \frac1\pi \int_0^\pi \tilde{g}(\phi)\, e^{-nu^2(\phi)/(4s_\gamma)} \;d\phi
\e{GG2}
$$
where
$$
\eqalign{
\tilde{g}(\phi)& =\frac{e^{\theta-r(\phi)}[\cos(\phi)\!+\!r'(\phi)\sin(\phi)] - 1}{ e^{2(\theta-r(\phi))} -2\cos(\phi) e^{\theta-r(\phi)} +1}\cr
r(\phi)& ={\rm asinh}(s_\gamma/\sinc(\phi))\;,\quad\sinc(\phi)=\sin(\phi)/\phi\cr
r'(\phi) & =  \frac{1/\phi-\cot(\phi)}{\sqrt{1+\sinc^2(\phi)/s_\gamma^2}}\cr
u(\phi) & =\sqrt{2h(\phi)}\cr
h(\phi) & = (1-\cos(\phi)) \cosh(r(\phi))+ \alpha(r(\phi))\;.
}
\e{GG4}
$$
\hfill~$\Box$\end{theorem}}

\Tom{Interestingly, in \e{GG2} we are dealing with a single-dimension integral. The fact that $n$ appears only at the exponent further makes the representation stable for numerical evaluation purposes. The representation of Theorem~\ref{teoTE0b} is also particularly interesting in light of the claimed equivalence between Shannon's and \ac{PPV} bounds, since it avoids the difficulties involved with the evaluation of two nested integrals (e.g., see \cite{wiechman:it08}).}

\subsection{\Tom{Asymptotic expansion}}

\Tom{An asymptotic expansion of the result straightforwardly follows from application of \cite[(4.3)]{Temme93}.}

\begin{theorem}\label{teoTE1}
An asymptotic expansion for $g(\gamma)$ which holds uniformly with respect to $n\in[1,\infty)$ can be expressed in the form
$$
g(\gamma)  = \frac1{2\sqrt{\pi}} \sum_{k=0}^\infty c_{2k} \frac{\Gamma(k+\frac12)}{\Gamma(\frac12)}\left(\frac{4s_\gamma}{n}\right)^{k+\frac12}\
\e{TE4}
$$
where real valued coefficients $c_{2k}$ are derived from Taylor expansion in $u=0$ of function $f(u)=i\sum_{k=0}^{\infty} c_{k} u^{k}$.\hfill~$\Box$
\end{theorem}

Observe that  the series expressing $g(\gamma)$ in \e{TE4} is a diverging series whose correct interpretation is that the remainder of the series after $k$ terms is, in Landau notation, $O(n^{-k-\frac12})$ as $n\rightarrow\infty$. This behavior is common to a number of widely used series expansions, e.g., the one for the error function \e{TE10c}. High precision can be obtained by appropriately limiting the series \Tom{to the index providing the smallest contribution.} Only the first few terms of this asymptotic expansion are needed, and, typically, in numerical evaluation it is often sufficient to limit the series to \Tom{a few entries} to get very high accuracy \Tom{(we will comment in later Section~III.E on methods to control the accuracy level)}. Note also that, for numerical stability, it is more appropriate to evaluate \ac{FA} and \ac{MD} probabilities \e{TE10} in logarithmic form, that is via the evaluation of $-\frac12 v(\gamma) +\frac1n\ln(\pm g(\gamma))$. 

Since \cite{Temme93} only provides expression for coefficients $c_0$ and $c_2$, we address in some detail the expression of the generic coefficient $c_{2k}$. These can be extrapolated by use of Taylor series expansions for nested functions. The result is made available in algorithmic form.

\begin{theorem}\label{teoTE2}
The coefficients $c_{2k}$ of Theorem~\ref{teoTE1} can be evaluated by the following procedure:
\begin{enumerate}
\item Define the real valued Taylor coefficients of function $\alpha(x)$ in $x=\gamma$ according to
$$
\alpha_n = \cases{ 0 & $n=0,1$\cr -c_\gamma/n! &  even $n$\cr -s_\gamma/n! &  odd $n\,$.\cr}
\e{NM4b}
$$
\item Derive the real valued Taylor coefficients of function $\nu(x)=\alpha(\ln(x))$ in $x=e^\gamma$ via
$$
\nu_n = \cases{0 & $n=0,1$\cr
\displaystyle \sum_{j=2}^n \alpha_j Q_{j,n} e^{-n\gamma} & otherwise,
}
\e{NM5b}
$$
where real valued constants $Q_{j,n}$ are iteratively defined by $Q_{n,n} = 1$ and 
$$
Q_{j,n} = \sum_{\ell=1}^{n-j} \frac{(\ell j-n+j+\ell)\,(-1)^\ell}{(n-j)(\ell+1)} Q_{j,n-\ell}\;.
\e{NM5}
$$
\item Derive the imaginary valued Taylor coefficients of function $\beta(x)=\sqrt{2\alpha(\ln(x))}$ in $x=e^\gamma$ via
$$
\beta_n = \cases{
0 & $n=0$\cr
-\sqrt{2\nu_2} & $n=1$\cr
\displaystyle \beta_1^{-1} \left(\nu_{n+1}- \fract12  \sum_{k=2}^{n-1} \beta_k\beta_{n+1-k}\right) & $n\ge2\,$.
}
\e{NM6}
$$
\item Derive Taylor coefficients of function $t(u)$ in $u=0$ via
$$
t_n = \cases{e^\gamma & $n=0$\cr
\beta_1^{-1} & $n=1$\cr
\displaystyle-\beta_1^{-n}\sum_{j=1}^{n-1} t_j P_{j,n} & $n\ge2\,$,}
\e{NM10}
$$
where coefficients $P_{j,n}$ are iteratively defined by $P_{n,n}=\beta_1^n$ and
$$
P_{j,n}  =  \sum_{\ell=1}^{n-j} \frac{(\ell j-n+j+\ell)\, \beta_{\ell+1}}{(n-j)\,\beta_1} P_{j,n-\ell} \;.
\e{NM12}
$$
Note that coefficients $t_{2n}$ (even entries) are real valued, while coefficients $t_{2n+1}$ (odd entries) are imaginary valued, and similarly is for coefficients $P_{j,n}$.
\item Derive Taylor coefficients of function $\xi(u)=1/(e^\theta-t(u))$ in $u=0$ using
$$
\xi_n = \cases{
(e^\theta-e^\gamma)^{-1} & $n=0$\cr
\displaystyle \sum_{j=1}^n (e^\theta-e^\gamma)^{-(j+1)}S_{j,n}& $n\ge1\,$,
}
\e{NM14}
$$
where $S_{n,n}=t_1^n$ and
$$
S_{j,n} = \sum_{\ell=1}^{n-j} \frac{(\ell j-n+j+\ell)\,t_{\ell+1}}{(n-j)\,t_1} S_{j,n-\ell}\;.
\e{NM16}
$$
Again, coefficients $\xi_{2n}$ are real valued, while coefficients $\xi_{2n+1}$ are imaginary valued.
\item Finally derive Taylor coefficients of $f(u)$ in $u=0$ using
$$
f_n = ic_n=  \sum_{k=0}^n (k+1) t_{k+1} \xi_{n-k}   \;.
\e{NM18}
$$
\end{enumerate}
\hfill~$\Box$
\end{theorem}

\subsection{\Tom{Remark on notation}} 

\Tom{In the following we will respectively denote with $P\sub{MD}^{(K)}$ and $P\sub{FA}^{(K)}$ the \ac{MD} and \ac{FA} asymptotic expressions obtained by limiting the series in \e{TE4} to $K$ contributions, i.e., by setting $k=0,\ldots,K-1$. The corresponding bounds on rate and error probability will be denoted, respectively, as $\overline{R}^{(K)}$ and $\underline{P}_e^{(K)}$.}

\subsection{\Tom{Compact expressions for the first two orders}}

\Tom{Theorem~\ref{teoTE2} is general, in the sense that it provides the coefficients expression $c_{2k}$ for any order $k$. We now look into the special case $k=0,1$ to compare to the results of \cite{Temme93}.} By applying the algorithm of Theorem~\ref{teoTE2} to the first orders\Tom{, and by using the shorthand notation 
$$
t_\gamma = \tanh(\gamma)\;,
\e{XZ2}
$$
}we then have
$$
\eqalign{
\alpha_2 & = -\fract12c_\gamma\;,\quad\alpha_3  = -\fract16s_\gamma\;,\quad \alpha_4  = -\fract1{24}c_\gamma\cr
\nu_2 & = -\fract12c_\gamma e^{-2\gamma}\;,\quad \nu_3 = \fract12c_\gamma(1-\fract13t_\gamma)e^{-3\gamma}\cr
\nu_4 & =  -\fract12c_\gamma(1-\fract12t_\gamma)e^{-4\gamma}\cr
\beta_1 & = -i\sqrt{c_\gamma}e^{-\gamma}\;,\quad \beta_2  = i\fract12\sqrt{c_\gamma}(1-\fract13t_\gamma)e^{-2\gamma}\cr
\beta_3 & =  -i\fract12\sqrt{c_\gamma}\Big[\fract34-\fract13t_\gamma-\fract1{36}t_\gamma^2\Big]e^{-3\gamma}\cr
t_1 & = i\frac{e^{\gamma}}{\sqrt{c_\gamma}}\;,\quad t_2  = \fract12\frac1{c_\gamma}(\fract13t_\gamma-1)e^{\gamma}\cr
t_3 & = -i\fract12\frac1{c_\gamma\sqrt{c_\gamma} }\Big[\fract14-\fract13t_\gamma+\fract5{36}t_\gamma^2\Big]e^{\gamma}\cr
\xi_0 & = (e^\theta-e^\gamma)^{-1}\;,\quad \xi_1  = i\frac{e^{\gamma}}{\sqrt{c_\gamma}}(e^\theta-e^\gamma)^{-2}\cr
\xi_2 & =t_2 (e^\theta-e^\gamma)^{-2}  -  \frac{1}{{c_\gamma}} e^{2\gamma} (e^\theta-e^\gamma)^{-3}\;,
}
\e{NM30}
$$
to obtain coefficients
$$
\eqalign{
c_0 & = \frac{1}{\sqrt{c_\gamma}(e^{\theta-\gamma}-1)} \cr
c_2 & = -i(t_1\xi_2 + 2t_2\xi_1+3t_3\xi_0) \;,
}
\e{NM34}
$$
which perfectly match to the expressions in \cite{Temme93}.

\Tom{Further insights can be obtained by assuming that, in the calculation of \ac{FA} and \ac{MD} probabilities, the two step functions in \e{TE10} are not active, that is 
$$
\fract12\ln\Big(1+\fract{\snr}{1+\snr}\Big) < \gamma < \fract12\ln(1+\snr) \;.
\e{NJ3}
$$
This corresponds to avoiding the calculation for \ac{FA} and \ac{MD} probabilities which are greater than $\frac12$, which is not a limitation in practice since it corresponds to the cases $P_e>\frac12$ (too large error probability) and $R<\frac1n$ (absence of reliable  communication, since the available symbols, $M=2^{nR}$, are less than two). With the above assumption the single-term and two-term asymptotic expansions take the form, respectively, of
$$
\eqalign{
\ln P\sub{MD}^{(1)} &  =  -\fract 12n\, v\sub{MD}(\gamma) -\fract12 \ln(n) +\ln (g_{0,\rm MD}(\gamma) )\cr
\ln P\sub{FA}^{(1)} &  =  -\fract 12n\, v\sub{FA}(\gamma) -\fract12 \ln(n) +\ln (-g_{0,\rm FA}(\gamma) ) \;,
}
\e{NJ4}
$$
and
$$
\eqalign{
\ln P\sub{MD}^{(2)} &  = \ln P\sub{MD}^{(1)} +\ln\Big(1-\fract1ng_{1,\rm MD}(\gamma)\Big) \cr
\ln P\sub{FA}^{(2)} &  =  \ln P\sub{FA}^{(1)} +\ln\Big(1-\fract1ng_{1,\rm FA}(\gamma)\Big) \;,
}
\e{NJ4b}
$$
where
$$
\eqalign{
g_0(\gamma) &= \sqrt{\frac{t_\gamma}{\pi}} \frac1{e^{\theta-\gamma}-1}\cr
g_1(\gamma) & = t_\gamma\left[ \frac{9\!-\!12t_\gamma\!+\!5t_\gamma^2}{12} + \frac{2\!+\!(3\!-\!t_\gamma)(e^{\theta-\gamma}\!-\!1)}{(e^{\theta-\gamma}-1)^2}\right]
}
\e{IL2bis}
$$
with $\theta$ taking a different value for \ac{MD} and \ac{FA} probabilities according to \e{TE4_v0}. Observe that, under \e{NJ3} it is $g_{0,\rm MD}(\gamma)\ge0$ for \ac{MD} probabilities, and $g_{0,\rm FA}(\gamma)\le0$ for \ac{FA} probabilities. Hence, the logarithmic expressions in \e{NJ4} always make sense. With a little effort it can be also verified that $g_1(\gamma)\ge0$, so that the two-terms asymptotic expansion always provides smaller probabilities. Incidentally, since it is $g_1(\gamma)=-2s_\gamma c_2/c_0$ by construction, then we are also guaranteed that ${\rm sign}(c_2)=-{\rm sign}(c_0)$.}

\subsection{\Tom{On the accuracy of the asymptotic expansion}}

\Tom{A certificate about the approximation error is available under the assumption that consecutive remainders in the series have opposite signs. In this case, Steffensen's error test \cite{steffensen2006interpolation} guarantees that the error is less than the first neglected term in the series, and has the same sign. There is a strong empirical evidence that this applies to the asymptotic expansion considered in this paper, at least in the \ac{SNR} ranges and for the parameters choices of interest. In these cases, although the general validity of the alternating rule on remainders' signs is difficult to demonstrate (this is due to the complexity of the functions involved), some strong results can be in any case given. Specifically, a method to identify wether the $K$-term asymptotic series is over/under estimating the true probability can be formalized as follows.
\begin{theorem}\label{theo98}
Under assumption \e{NJ3}, a sufficient condition for the validity of bounds
$$
\eqalign{
P\sub{MD}^{(L)} &\le P\sub{MD} \le P\sub{MD}^{(U)}\cr
P\sub{FA}^{(L)} &\le P\sub{FA} \le P\sub{FA}^{(U)}\;,
}
\e{KL4}
$$
is that 
$$
1 - \sum_{k=1}^{L-1} \frac{c_{2k}}{c_0} u^{2k}(\phi) \le \frac{c(\phi)}{c_0} \le 1 - \sum_{k=1}^{U-1} \frac{c_{2k}}{c_0} u^{2k}(\phi)
\e{KL2}
$$
holds in $\phi\in[0,\pi)$ for both \ac{MD} and \ac{FA} probabilities, where
$$
\eqalign{
c(\phi)  & = \tilde{g}(\phi)\, u(\phi)/h'(\phi)\cr
h'(\phi) & = \sin(\phi)  \cosh(r(\phi)) \left[1+ (r'(\phi))^2  \right] 
}
\e{NJ12}
$$
and where the remaining functions were defined in \e{GG4}.\hfill~$\Box$
\end{theorem}}

\Tom{Applicability of Theorem~\ref{theo98} to consecutive series expansion, $L=U+1$ or $L=U-1$, states that the next term in the series limits the residual error in the asymptotic expansion (as it is for the $Q$-function series \e{TE10c}), and can be used to identify the error magnitude (and sign). Moreover, since both \ac{MD} and \ac{FA} probabilities \e{MO30} are monotone in $\lambda'$ (or $\gamma$), another straightforward consequence of  the applicability of Theorem~\ref{theo98} is the following result, 
\begin{corollary}\label{theo98b}
Applicability of Theorem~\ref{theo98} ensures that bounds 
$$
\eqalign{
\overline{R}^{(U)} & \le \overline{R}\le \overline{R}^{(L)}\cr
\underline{P}_e^{(L)} & \le \underline{P}_e \le  \underline{P}_e^{(U)}\cr
}
\e{KL4b}
$$
apply, respectively, to \ac{PPV} bounds on rate and error probability. \hfill~$\Box$
\end{corollary}
Note that the relation in the first of \e{KL4b} is reversed with respect to the findings of Theorem~\ref{theo98} due to the fact that the upper bound on rate implies a $-$ sign.}

\Tom{Although \e{KL2} does not hold in general for any parameters choice, there is numerical evidence that it holds in the cases of practical interest for application (e.g., for all the numerical examples developed in \sect{EX}). In addition, the property certainly holds with choice $U=1$, as stated by the following theorem.}

\Tom{\begin{theorem}\label{theo99}
Under \e{NJ3}, the single-term asymptotic expansion always provides an upper bound to both \ac{MD} and \ac{FA} probabilities, that is $P\sub{MD} \le P\sub{MD}^{(1)}$ and $P\sub{FA} \le P\sub{FA}^{(1)}$. As a consequence, the following bounds can be established
$$
\overline{R}^{(1)} \le \overline{R}\;,\qquad \underline{P}_e \le  \underline{P}_e^{(1)}\;.
\e{KL4bbb}
$$
\hfill~$\Box$
\end{theorem}}\Fig[!t]{FB2}\Fig[!t]{FB6}

\Section[EX]{\Tom{Application} Examples}

\subsection{\Tom{Numerical evaluation of the converse bound}}

A thorough overview on the converse bound is given in the graphs of \fig{FB2} and \fig{FB6}, \Tom{showing the spectral efficiency upper limit $\overline{\rho}=2\overline{R}$ for $P_e=10^{-5}$ and for a wide range of values of $n$. The spectral efficiency is plotted in solid lines versus the \ac{SNR} per symbol ($\snr$) in \fig{FB2}, and versus the \ac{SNR} per bit ($E_b/N_0=\snr/\rho$) in \fig{FB6}. Note that, in \fig{FB2} the bound is set to zero below a certain \ac{SNR} (for $n=10$ and $20$). This corresponds to the region where $\overline{R}<\frac1n$, implying that $M<2$ transmission codewords are available, which denotes the absence of reliable communication (i.e., a single codeword $M=1$).}

\Tom{Curves are obtained} by use of the numerical method developed in Theorems~\ref{teoTE0}-\ref{teoTE2}. The method is implemented in MatLab, using up to $21$ active coefficients for the series of $g$, \Tom{the series being truncated in correspondence to the smallest contribution to obtain the best accuracy. The assumption of Theorem~\ref{theo98} was always numerically verified to hold, hence a guarantee on precision is available. Precision was found to be well below $1\%$ for rates $\overline{R}>\frac3n$, which satisfactorily covers the region of practical interest, since its is neglecting only those cases where less than $M=2^3=8$ codewords are available. For rates $\overline{R}<\frac3n$, which are very close to the absence-of-reliable-communication limit, the integral form was used to get reliable results.}

\Tom{The range of applicability of the first and second-order series expansion \e{NJ4} and \e{NJ4b} is illustrated in \fig{FB6},  showing (in dash-dotted lines) the lower bound $\overline{\rho}^{(1)}=2\overline{R}^{(1)}$ and the two-term approximation $\overline{\rho}^{(2)}=2\overline{R}^{(2)}$ in the region where their disagreement is limited, which is the region where $\overline{R}>\frac4n$. We numerically verified that the assumption of Theorem~\ref{theo98} holds in this region with $L=2$, so that \e{NJ4} and \e{NJ4b} are guaranteed to be reliable approximations of the true value of $\overline{\rho}$, and $\overline{R}^{(2)}$ is also guaranteed to be an upper bound. In this context, quantity $g_1(\gamma)/n$ is a measure of the (normalized) impact of the residual error.}

\subsection{\Tom{Asymptotic behavior for $n\rightarrow\infty$}}

\Tom{We observe from \fig{FB2} that} the bound increases with $n$, until it ultimately touches the Shannon's limit, $\rho=\log_2(1+\snr)$, for $n\rightarrow\infty$. \Tom{This can be easily verified analytically.} As a matter of fact, at the limit $n\rightarrow\infty$ the constraint $P\sub{MD}=\epsilon$ provides $v\sub{MD}(\gamma)=0$, which implies $\gamma=\theta\sub{MD}(\gamma)$, and so 
$$
\lim_{n\rightarrow\infty} \gamma = \overline{\gamma}= \fract12\ln (1+\snr) \;. 
\e{OK2}
$$
Hence, by substitution, it is $\theta\sub{FA}(\overline{\gamma})= -\frac12\ln (1+\snr)=-\overline{\gamma}$, and $v\sub{FA}(\overline{\gamma})=\ln (1+\snr)$. Then we have
$$
\lim_{n\rightarrow\infty} \overline{R} = C\;,
\e{NM36}
$$
independently of the value of $\epsilon$ and $\snr$, but provided they are finite.

\subsection{\Tom{Optimal parameters choice in the power limited regime}}

\Tom{Some further insights can be inferred from \fig{FB6}.} The ultimate limit in figure is $E_b/N_0=\ln2=-1.59\,$dB, reached for $n\rightarrow\infty$ and $\overline{\rho}\rightarrow0$. Note that, closing the gap to this ultimate limit requires very long codes: a $1.2\,$dB gap is experienced with $n=10^4$, narrowing to $0.6\,$dB with $n=10^5$, and to $0.3\,$dB with $n=10^6$. Moreover, each block-length is $E_b/N_0$ optimal at a different spectral efficiency, which suggests that the two should be matched for optimal performance. All these results confirm the inner difficulty in optimizing codes performance in the power limited regime. \Tom{Incidentally, these results are confirmed by \fig{FB26}\Fig[!t]{FB26} where the converse bound is plotted against the so called $\kappa\beta$ achievability bound (see \cite[Theorems~25,42,43]{polyanskiy2010channel})
$$
\eqalign{
\underline{R}_{\kappa\beta} = & \max_\lambda \;\frac1n\log_2{\rm erf}\left(\frac{\sqrt{1\!+\!2\snr}}{1\!+\!\snr} \alpha\right)-\frac1n\log_2 P\sub{FA}(\lambda)\cr
& \hbox{subject to }  P\sub{MD}(\lambda) = P_e-{\rm erf}(\alpha)\cr
&\phantom{\hbox{subject to }} 0< \alpha< {\rm erf}^{-1}(P_e)\;,
}
\e{LI4}
$$
where ${\rm erf}$ is the error function, and against the $O(n^{-1})$ normal approximation (see \cite[Theorem~54]{polyanskiy2010channel}, and \cite{tan2015it})
$$
{R}\sub{NA} = C - \log_2(e)Q^{-1}(P_e)\sqrt{\frac{\snr (2+\snr)}{2n(1+\snr)^2}}  + \frac{\log_2(n)}{2n} \;.
\e{LI2}
$$
Observe from \fig{FB26} that the three curves have a similar behavior, their difference vanishing at large block-lengths. Also observe that the normal approximation is not always a reliable approximation of the \ac{PPV} bound, especially in the region of small spectral efficiency, and in connection with small block-lengths $n$.}

\subsection{\Tom{Further relations with the literature}}

\Tom{Evaluation of the bound using the settings of \cite{polyanskiy2010channel} is shown in \fig{FB30} and \fig{FB31}.\Fig[!t]{FB30}\Fig[!t]{FB31} The plots fully adhere to \cite[Fig.~6-7]{polyanskiy2010channel}, with a fundamental difference:  an explicit expression is available, together with a guarantee (given by Theorem~\ref{theo98}) on the fact that we are calculating the true upper bound $\overline{R}$. Numerical effectiveness can be appreciated by comparing  \fig{FB31} to \cite[Fig.~7]{polyanskiy2010channel}, and by observing that: 1) we are obtaining a tighter result $\overline{R}$ for low values of $n$; and 2) the proposed numerical approach does not  show any numerical inconsistency as $n$ grows.}

\subsection{\Tom{Deriving the normal approximation}}

\Tom{We observe that the proposed expansion allows an easy derivation of the normal approximation \e{LI2}. The derivation first requires identifying an asymptotic approximation of the solution to $P\sub{MD}(\gamma)=P_e$. To this aim, by assuming that \e{NJ3} applies, we write the \ac{MD} probability \e{TE10} in the form
$$
\eqalign{
\ln P\sub{MD}(\gamma)  
& = \frac{n\alpha(\theta\sub{MD}(\gamma))}{2s_\gamma} -\ln\left(\frac{e^{\theta\sub{MD}(\gamma)-\gamma}\!-\!1}{\sqrt{t_\gamma/n \pi}}\right)\!+\! O(1/n)\cr
&\hspace*{-2mm} =  \frac{n\alpha(\theta\sub{MD}(\gamma))}{2s_\gamma} +\ln q\left(\frac{e^{\theta\sub{MD}(\gamma)-\gamma}\!-\!1}{\sqrt{2t_\gamma/n}}\right)\!+\! O(1/n)\cr
}
\e{XC2}
$$
where we exploited \e{TE10c}. The result in \e{XC2} is well defined since, as explained in \cite{Temme93}, the error function serie given by $q(\cdot)$ in \e{TE10c} is dealing with the presence of a pole in the Laplace domain. By then Taylor expanding (up to the first non null order) the functions in \e{XC2} in $\gamma=\overline{\gamma}$, which is the limit value according to \e{OK2}, we obtain
$$
\eqalign{
\ln P\sub{MD}(\gamma)  
& = \ln Q\left(x \sqrt{1+O(\overline{\gamma}-\gamma)}\;\right)  + O(1/n)\;,
}
\e{XC2a}
$$
where
$$
x = \sqrt{n}\,\sqrt{\frac{2(1+\snr)^2(2+\snr)}{\snr^3}}\,(\overline{\gamma}-\gamma)\;,
\e{XC2b}
$$
and where we exploited $\ln Q(x) = -\frac12x^2 + \ln q(x)$. Note from \e{XC2a} that $x$ in the limit tends to the non null constant value $Q^{-1}(P_e)$, and therefore the contribution $O(\overline{\gamma}-\gamma)$ can be replaced by $O(1/\sqrt{n})$. By inverting $P\sub{MD}(\gamma)=P_e$ using \e{XC2a} we then obtain
$$
\gamma = \overline{\gamma} - Q^{-1}(P_e) \sqrt{\frac{\snr^3}{2n(1+\snr)^2(2+\snr)}} + O(1/n)\;,
\e{XC6}
$$
where the correctness of the big-O notation is ensured by standard series expansions for $Q^{-1}$. By now switching to the \ac{FA} probability \e{TE10}, which does not suffer from the presence of a pole in the Laplace domain at the limit, by Taylor expansion at $\gamma=\overline{\gamma}$ we have
$$
\eqalign{
-\frac1n\ln P\sub{FA} & = -\frac{\alpha(\theta\sub{FA}(\gamma))}{2s_\gamma} + \frac{\log(n)}{2n} + O(1/n)\cr
 & = \overline{\gamma} - \frac{2+\snr}{\snr} (\overline{\gamma}-\gamma) + \frac{\log(n)}{2n}+ O(1/n)\;.
}
\e{UJ2}
$$
Substitution of \e{XC6} in \e{UJ2} finally provides the normal approximation \e{LI2}. The result identified by \e{TE10} is therefore consistent with the findings of the literature.}

\Tom{Incidentally, the above stated relation between the single-term approximation $\overline{R}^{(1)}$ and the normal approximation $R\sub{NA}$,  further explains why it is empirically observed that $R\sub{NA}\le\overline{R}$, the justification being given by Theorem~\ref{theo99}.}

\subsection{\Tom{Excess power}}

Some further useful insights are given in \fig{FB4}\Fig[!t]{FB4} which shows the excess power, $\Delta\snr$, over the one predicted by channel capacity to achieve the same spectral efficiency. This corresponds, in \fig{FB2}, to the value of the horizontal gap with respect to the Shannon bound, that is to
$$
[\Delta\snr]\sub{dB} = 10\log_{10}\snr - 10\log_{10}\Big(2^{2\overline{R}(\snr)}-1\Big)\;.
\e{NM39}
$$
Interestingly, note how the gap saturates for large \ac{SNR}, and that small gaps require very large block-lengths. For example, at $\snr=0\,$dB a $0.1\,$dB excess power is obtained with block-lengths between $10^5$ and $10^6$. \Tom{Interestingly, a similar gap is experienced with binary \ac{LDPC} codes \cite{richardson2001design}, but we warn the reader that two situations are not fully comparable since the considered \acp{LDPC} are built on binary codewords, i.e., on a subset of $\C F$ in \e{MO28}.}

\subsection{\Tom{Optimal parameters choice in the bandwidth limited regime}}

The behavior of the excess power $\Delta\snr$  \e{NM39} of \fig{FB4} for $\snr\rightarrow\infty$ \Tom{(the bandwidth limited regime according to \cite{forney:it98})} can be captured in a similar way, but the derivation is more involved, and it is therefore presented in the form of a theorem, whose proof is available in the Appendix. 

\begin{theorem}\label{leTEO4R}
The asymptotic behavior of the converse bound $\overline{R}$ and of the excess power $\Delta\snr$ for $\snr\rightarrow\infty$, and for fixed packet error rate $P_e$ and block-length $n$, is given by
$$
\eqalign{
\overline{R} & = \fract12\log_2(\snr) -\left[\frac1n\log_2q\Big(\sqrt{n/2}\Big)+\fract12\log_2(\lambda')\right]\cr
[\Delta\snr]\sub{dB} & = \frac{20}n\log_{10}q\Big(\sqrt{n/2}\Big)+10\log_{10}(\lambda' )\;,
}
\e{AS20}
$$
where $q$ was defined in \e{TE10c}, and $\lambda'$ is the solution to equation
$$
q\Big(\sqrt{n \fract12(\lambda'-1)^2}\Big)\,e^{ -\frac12n(\lambda'-1-\ln (\lambda'))} = P_e\;.
\e{AS22}
$$
\hfill~$\Box$
\end{theorem}

The result for $P_e=10^{-5}$ is shown in \fig{FB4B}\Fig[t]{FB4B}. Note the perfect correspondence with \fig{FB4}, and how the behavior at large block-lengths $n$ is approximately linear in logarithmic scale. Specifically, for large $n$ the second order approximation $\lambda'-1-\ln (\lambda')\simeq \frac12(\lambda'-1)^2$ holds in \e{AS22}, and therefore we have $\Delta\snr\sim 10\log_{10}\lambda'$, with $\lambda'=1+\sqrt{2x/n}$, and $x$ the solution to $Q(\sqrt{x})=q(\sqrt{x})e^{-x/2}=P_e$. The linear approximation follows as 
$$
[\Delta\snr]\sub{dB} \sim 10\log_{10}e \cdot\sqrt{\frac2n} \,Q^{-1}(P_e)\;,
\e{IL0}
$$ 
and it is shown in dashed lines in \fig{FB4B}. The corresponding approximation on $\overline{R}$ provides
$$
\overline{R} \sim C - \log_2(e)\, Q^{-1}(P_e)\, \sqrt{\frac{1}{2n}}\;,
\e{IL0b}
$$
which is in perfect agreement with the \Tom{normal approximation \e{LI2} for $\snr\rightarrow\infty$ (see also the limit expression in \cite{Ingber:it13})}.

\subsection{Packet error rate perspective}

A different view is finally provided by \fig{FB8}, which shows packet error rate performance versus \ac{SNR} for rate $R=\frac12$, making use of \e{MO27}. The converse bound $\underline{P}_e$ is to be intended, in this case, as a lower bound to the best achievable performance. Note from the figure how for $\snr>0$, the rate achieving \ac{SNR}, the performance improves with block-length $n$. For completeness, the (achievable) bound for $n=1$, namely $P_e=Q(\sqrt{\snr})$, is also shown. The upper bound $\underline{P}_e^{(1)}$ is illustrated in dash-dotted lines in  \fig{FB8}, and is shown to provide a very good approximation down to very low values of $n$.\Fig[!t]{FB8} Comparison with the normal approximation in \e{LI2} is illustrated in \fig{FB8B}\Fig[t]{FB8B}.

\section{Conclusions}

In this paper we provided a means to reliably evaluate the Polyanskiy-Poor-Verd\'u converse bound in \ac{AWGN}. The \Tom{proposal consist of a (single) integral form, and of an asymptotic expansion which allows evaluating the bound with great precision via simple expressions. A method to control the accuracy of the result is also provided. Comparison with the widely used normal approximation suggests that the proposed solution is of interest especially for medium/low block-lengths ($n<1000$).} The importance of the \Tom{contribution} should be also read in connection with the rapidly increasing interest on simple but meaningful descriptions of the communication performance in the finite block-length regime, e.g.,  to be used in the optimization process of MAC (medium access control) or higher layers \Tom{in future M2M (machine to machine) communication scenarios. }

\appendix

\begin{IEEEproof}[Proof of Theorem~\ref{teoPOLawgn}]
For the sake of clarity, the result is obtained by application of the Neyman-Pearson test to the \ac{AWGN} case, rather than from reinterpretation of \cite[Theorem~40]{Polyanskiy10}. In the \ac{AWGN} case we can write the Neyman-Pearson log-likelihood ratio \e{MO10} in the form 
$$
\eqalign{
\Lambda'(\B x,\B y) 
	 & = \frac1n\left\|\frac{\B y}{\sigma_w}-\frac{\B x}{\sigma_w}\frac{1+\snr}{\snr}\right\|^2 \cr
	 & = \frac1n (1+\snr) \left\|\frac{\B y}{\sigma_y}-\frac{\B x}{\sigma_x}\sqrt{\frac{1+\snr}{\snr}}\right\|^2_{\phantom{\big|}},
}
\e{II20}
$$
which is equivalent to \e{MO10} up to a negative multiplication factor and an addition factor (the relation is equivalent to \e{LL2} given for the corresponding decision thresholds $\lambda$ and $\lambda'$). Hence, the correct Neyman-Pearson test formulation is to choose $\C H_1$ if $\Lambda'(\B x,\B y) \le \lambda'$, and $\C H_0$ otherwise. In this context for the \ac{MD} probability we have
$$
\eqalign{
& P\sub{MD}(\B x,\lambda)  \cr
 & = \P{ \left.\left\|\frac{\B y}{\sigma_w}-\frac{\B x}{\sigma_w}\frac{1+\snr}{\snr}\right\|^2>n\lambda'\right|
 	 \B y\sim \C N\big(\B x,\B I_n\sigma_w^2\big)}\cr
 & = \P{ \left\|\B s\right\|^2>n\lambda'\left| \B s\sim \C N\left(-\frac{\B x}{\sigma_w\snr},\B I_n\right)\right.}\cr
 & = \P{ x >n\lambda'\left| x\sim \chi\left(n,\left\|\frac{\B x}{\sigma_w\snr}\right\|^2=\frac n\snr\right)\right.}\cr
}
\e{NG8}
$$
providing the first of \e{MO30}. For the \ac{FA} probability we have
$$
\eqalign{
& P\sub{FA}(\B x,\lambda)\cr
 & = \P{ \left.\left\|\frac{\B y}{\sigma_y}-\frac{\B x}{\sigma_x}\sqrt{\frac{1+\snr}{\snr}}\right\|^2\le\frac{n\lambda'}{1+\snr}\right|
 	 \B y\sim \C N\big(\B 0,\B I_n\sigma_y^2\big)}\cr
 & = \P{ \left.\|\B s\|^2\le\frac{n\lambda'}{1+\snr}\right| \B s\sim \C N\left(-\frac{\B x}{\sigma_x}\sqrt{\frac{1+\snr}{\snr}},\B I_n\right)}\cr
 & = \P{\left. x \le\frac{n\lambda'}{1+\snr}\right| x\sim \chi\left(n,n\frac{1+\snr}{\snr}\right)}\;,
}
\e{NG10}
$$
providing the second of \e{MO30}. Since both $P\sub{MD}(\lambda|\B x)$ and $P\sub{FA}(\lambda|\B x) $ are independent of the value of $\B x$, they are equal to their average counterparts, and the theorem is proved.
\end{IEEEproof}

\begin{IEEEproof}[Proof of Theorem~\ref{achieveandconverse}] 
Consider an encoding/decoding procedure ensuring $P_e$ \Tom{(average probability of error)} with equally likely messages $p_W(i)=\frac1M$. Use this setup by assuming that the channel now exhibits transition probabilities $p_{y|x}$ under $\C H_1$, and $q_y$ under $\C H_0$. Use the above to build a (suboptimal) binary hypothesis test that decides between $\C H_1$ and $\C H_0$ by observing $(\B x,\B y)$. The outcome is built in such a way that $\C H_1$ is selected if $\widehat{W}=W$. \ac{MD} and \ac{FA} probabilities for this test are, respectively, $P\sub{MD} = {\rm P}[\widehat{W}\neq W| \C H_1] = P_e$, and $P\sub{FA}  =  {\rm P}[\widehat{W}= W| \C H_0]$ given by
$$
\eqalign{ 
P\sub{FA}  
	& = \sum_{i=1}^M\sum_{\Bs a\in\C C}\sum_{\Bs b\in\C A_y^n}  \P{\widehat{W}=i,\B y=\B b,\B x=\B a, W=i\Big|\C H_0} \cr
	& = \sum_{i=1}^M\sum_{\Bs a\in\C C}\sum_{\Bs b\in\C A_y^n} p_{\widehat{W}|y}(i|\B b)\cdot q_y(\B b) \cdot \delta_{\Bs a,\Bs c_i}\cdot \frac1M= \frac1M\;.
}
\e{II2}
$$
\Tom{Observe that $P\sub{MD}$ depends upon the chosen encoder/decoder procedure (through $P_e$), while $P\sub{FA}$ is independent of it.} We then exploit the Neyman-Pearson lemma (e.g., see \cite{kay:93}) stating that: between all (possibly randomized) binary hypotheses tests on $\C H_0$ and $\C H_1$ that guarantee a given \ac{MD} probability, the Neyman-Pearson test is the one providing the smallest \ac{FA} probability. Hence, by selecting a Neyman-Pearson test with $P\sub{MD}(\lambda)=P_e$,  we obtain $P\sub{FA}(\lambda)\le \frac1M$, which proves the theorem \Tom{for average probability of error}.

\Tom{For maximal probability of error  we can use the same argument to write $P\sub{MD}(\lambda)\le P_e$ since average error probability is by definition smaller than or equal to maximal error probability. The inequality $P\sub{FA}(\lambda)\le \frac1M$ is also valid. The resulting bound is identical to the one of the average probability.}
\end{IEEEproof}

\begin{IEEEproof}[Proof of Theorem~\ref{teoTE0}]
Preliminarily note that, by exploiting standard results on \acp{CDF} of non-central chi-square random variables \cite{proakis:95}, \e{MO30} can be written in the form
$$
\eqalign{
P\sub{MD} & = Q_{\frac n2}\left(\sqrt{\frac{n}\snr},\sqrt{n\lambda'}\right)\cr
P\sub{FA} & = 1-Q_{\frac n2}\left(\sqrt{n\frac{1+\snr}{\snr}},\sqrt{\frac{n\lambda'}{1+\snr}}\right)\;,
}
\e{PT2}
$$
where
$$
\eqalign{
Q_\mu(a,b) & = \int_b^\infty \frac{z^\mu}{a^{\mu-1}} e^{-\frac12(z^2+a^2)} I_{\mu-1}(az) dz\cr
& = \int_{y}^\infty \left(\frac{z}{x}\right)^{\frac{\mu-1}2} e^{-z-x} I_{\mu-1}(2\sqrt{xz}) dz
}
\e{PT4}
$$
is the Marcum Q function expressed in the standard from (first line), and in the alternative form used by \cite{Temme93} (second line), where $x=\frac{a^2}2$ and $y=\frac{b^2}2$. In our context, parameters of the alternative form assume values $\mu=\frac n2$ and 
$$
x = \frac \mu\snr \cdot \cases{ 1 & MD \cr 1+\snr & FA}\;,\quad
y = \mu\lambda'  \cdot \cases{ 1 &MD \cr \frac1{1+\snr}& FA.}
\e{PT8}
$$
We finally apply \cite[(4.2)]{Temme93} to our specfic setting to obtain \e{TE10}. Note that in the notation of \cite{Temme93} we have $\xi=2\sqrt{xy}=n\sqrt{\lambda'/\snr}$, $\beta=n/(2\xi)= \frac12\sqrt{\snr/\lambda'}$, and $\sinh(\gamma)=\beta$ with $\gamma>0$. \Tom{Note also that function $\alpha$ in \e{TE4_v0} is different from the $\alpha$ function used in \cite{Temme93}, although closely related to it.} The value of $\lambda'$ in \e{TE10a} is a direct consequence of this result. Also the value of $\theta=\frac12\ln(y/x)$ in \e{TE4_v0} is a direct consequence of its definition, and of the property $\frac12\ln(\snr\lambda')=\ln(\snr)-\ln(2\sinh(\gamma))$. Value $s$ in \e{TE4_v0} corresponds to the sign of $u_0$ in \cite{Temme93}, and $v$ corresponds to $u_0^2\xi/n=\frac12u_0^2/\beta$. Differently from \cite[(4.2)]{Temme93}, in our formulation we avoided the erfc term by deleting the \Tom{contribution} $1/(u-iu_0)$ in the definition of $f(u)$, and by appropriately inserting the unit step function in \e{TE10}. \Tom{Finally, note that the integral defining $g(\gamma)$ is limited to the positive axis thanks to the fact that, by construction, $\Re[f(u)]$ is odd, and $\Im[f(u)]$ is even.} 
\end{IEEEproof}

\Tom{\begin{IEEEproof}[Proof of Theorem~\ref{teoTE0b}]
The result in \e{GG2} is obtained by working on an alternative expression for $u$, which exploits the way the integral is approached in \cite{Temme93}. We therefore use $r(\phi)$ in \e{GG4} to generate $u\in[0,\infty)$ according to the map $u(\phi)=t^{-1}(e^{r(\phi)+i\phi})=\sqrt{2\alpha(r(\phi)+i\phi)}$ for $\phi\in[0,\pi)$, which provides the expression in \e{GG4}.
The corresponding value of $t$ is $t(u(\phi)) = e^{r(\phi)+i\phi}$. With a little effort we can then write 
$$
\eqalign{
c(\phi)  & = \Im[f(u(\phi))]\cr & =  \Im\left[\frac{1}{e^{\theta-r(\phi)-i\phi}-1} \cdot \frac{u(\phi)}{\alpha'(r(\phi)+i\phi)}\right]\cr
 & = \underbrace{\Im\left[\frac{r'(\phi)+i}{e^{\theta-r(\phi)-i\phi}-1}\right]}_{\tilde{g}(\phi)} \cdot
 	\underbrace{\frac{\sqrt{2h(\phi)}}{h'(\phi)}}_{1/u'(\phi)}
} 
\e{GD2}
$$
where the contribution of  $\tilde{g}(\phi)$ is expressed in a more direct form in \e{GG4}. The integral form \e{GG2} is derived from \e{TE4_v0} and \e{GD2} by a change of variable. Incidentally, note also that function $h(\phi)$ is real valued, increasing, and positive by construction. It is in fact built real and positive in order to identify the steepest descent path.
\end{IEEEproof}}

\begin{IEEEproof}[Proof of Theorem~\ref{teoTE2}] 1) Coefficients \e{NM4b} are a straightforward consequence of the Taylor expansion of the hyperbolic cosine. 2) We exploit Taylor expansion of $\ln(x)$ around $x=e^\gamma$, namely
$$
\ln(x) = \gamma + \sum_{n=1}^\infty \frac{(-1)^{n+1}}{n} \left(\frac{x-e^\gamma}{e^\gamma}\right)^n\;.
\e{II30}
$$
Then, from identity (e.g., see \cite{Itskov12})
$$
\left(\sum_{n=1}^\infty \frac{(-1)^{n+1}}{n}\epsilon^n \right)^j = \sum_{k=j}^{\infty} Q_{j,k} \epsilon^k\;,
\e{II32}
$$
where coefficients are defined as in \e{NM5}, we have
$$
\nu(x) = \sum_{j=2}^\infty \alpha_j \sum_{k=j}^{\infty} Q_{j,k} \left(\frac{x-e^\gamma}{e^\gamma}\right)^k\;,
\e{II34}
$$
which proves \e{NM5b} by swapping the summations order. 3) From equivalence $2\nu(x) = \beta^2(x)$ we can write 
$$
2\sum_{n=0}^\infty \nu_n (x-e^\gamma)^n = \sum_{n=0}^\infty \left(\sum_{k=0}^{n}  \beta_k\beta_{n-k}\right)  (x-e^\gamma)^n\;,
\e{II36}
$$
providing the equivalences
$$
\eqalign{
\beta_0^2 & =  0 \cr 
2\beta_0\beta_1 & = 0 \cr 
2\beta_0\beta_2 + \beta_1^2 & = 2\alpha_2 \cr
2\beta_0\beta_n + 2\beta_1\beta_{n-1} + \sum_{k=2}^{n-2} \beta_k\beta_{n-k}& = 2\alpha_n\;.
}
\e{II38}
$$
These can be exploited to obtain \e{NM6}. Note that also $\beta_1=\sqrt{2\nu_2}$ is a viable choice for inversion, the difference being a sign inversion on coefficients $\beta_n$, but this provides an unwanted sign inversion in the evaluation of $t(u)$. 4) The result is obtained by exploiting the method of \cite{Itskov12} for the inverse function $\beta^{-1}(x)$. The correctness of the result can be checked by recalling from \cite{Temme93} that couples $(u,t(u))$ are obtained from $t(u)= e^{r(\phi)+i\phi}$ and $u(\phi)$ as defined in \e{GG4}. 5) We exploit the same method used in 2). Specifically, \e{NM14} follows from Taylor expansion  of
$$
\frac1{e^\theta-x} = \sum_{n=0}^{\infty} \frac1{(e^\theta-e^\gamma)^{n+1}} (x-e^\gamma)^n\;,
\e{II42}
$$
and from identity
$$
(f(u)-e^\gamma)^j = \left(\sum_{n=1}^\infty t_n u^n\right)^j =   \sum_{k=j}^\infty S_{j,k} u^k\;,
\e{II44}
$$
with $S_{j,k}$ as in \e{NM16}. 6) The result can be inferred by expressing product $t'(u)\xi(u)$ in the form
$$
t'(u)\xi(u) = \sum_{k,\ell=0}^\infty  (k+1)t_{k+1} \xi_\ell u^{k+\ell}\;,
\e{II46}
$$
and by then rearranging the summation as a summation in $k$ and $n=k+\ell$.
\end{IEEEproof}

\Tom{\begin{IEEEproof}[Proof of Theorem~\ref{theo98}]
We prove the upper bound, since the lower bound can be derived in a perfectly equivalent way. Define $c(u)= \Im[f(u)]$. For \ac{MD} probabilities it is $s\sub{MD}(\gamma)<0$, hence $c(u)\le  \sum_{k=0}^{U-1} c_{2k}u^{2k}$ is a sufficient condition for ensuring an upper bound on $P\sub{MD}$.  A similar expression is obtained for \ac{FA} probabilities, but in this case there is a $-$ sign involved in \e{TE10}, which turns the condition into $c(u)\ge \sum_{k=0}^{U-1} c_{2k}u^{2k}$. Given that, under \e{NJ3}, $c_0$ has positive sign with \ac{MD} probabilities, and negative sign with \ac{FA} probabilities, then the condition can be written in the form
$$
\frac{c(u)}{c_0} \le 1 +  \sum_{k=1}^{U-1} \frac{c_{2k}}{c_0} u^{2k}
\e{NJ11}
$$
for both \ac{MD} and \ac{FA} probabilities. In order to proceed, it is useful to work on $u(\phi)$, which provides \e{KL2} and \e{NJ12} because of \e{GD2}.
\end{IEEEproof}}

\Tom{\begin{IEEEproof}[Proof of Theorem~\ref{theo99}]
We want to prove that $c(\phi)/c_0\le1$ for both \ac{FA} and \ac{MD} probabilities, in which case the assumptions of Theorem~\ref{theo98} hold, and the theorem is proved. To do so, we first express the fraction $c(\phi)/c_0$ as the product $b_1(\phi)b_2(\phi)$ where
$$
\eqalign{
b_1(\phi) & = \frac{\sqrt{2h(\phi)c_\gamma}}{h'(\phi)} 
\cr
b_2(\phi) & = \frac{(e^{\theta-\gamma}-1) \Big(e^{\theta-r(\phi)}[\cos(\phi)\!+\!r'(\phi)\sin(\phi)] - 1\Big)}{ e^{2(\theta-r(\phi))} -2\cos(\phi) e^{\theta-r(\phi)} +1} \;.
}
\e{DM2}
$$
Note that, thanks to the fact that $-\cos(\phi)\ge-1$, the denominator of $b_2(\phi)$ is a positive quantity. Also all the factors in $b_1(\phi)$ are positive by construction. We then identify an upper bound for both $b_1(\phi)$ and $b_2(\phi)$, by assuming $\phi\in[0,\pi)$. By exploiting the fact that $r(\phi)\ge\gamma$, and that $\alpha(x)$ is negative for $x\ge\gamma$ (see the Taylor expansion \e{NM4b}), we have $h(\phi)\le(1-\cos(\phi))\cosh(r(\phi))$. Hence, by substitution in the first of \e{DM2} we obtain 
$$
b_1(\phi) \le \frac{\sinc(\fract12\phi)}{\sinc(\phi)}\frac1{1+(r'(\phi))^2} \sqrt{\frac{c_\gamma}{\cosh(r(\phi))}}\;.
\e{DM6}
$$
Instead, the upper bound for $b_2(\phi)$ requires distinguishing between \ac{MD} and \ac{FA} probabilities, and to exploit the property
$$
1-\left(\frac{c_\gamma-s_\gamma}{c_\gamma+s_\gamma}\right) (1-\sinc^2(\phi)) \le e^{r(\phi)-\gamma}\sinc(\phi) \le 1\;,
\e{DM8}
$$ 
which can be obtained from the equivalence
$$
e^{r(\phi)-\gamma}\sinc(\phi) = \frac{s_\gamma+\sqrt{s_\gamma^2+\sinc^2(\phi)}}{s_\gamma+c_\gamma}
\e{DM10}
$$
by recalling that $0\le\sinc(\phi)\le1$ holds for $\phi\in[0,\pi)$. Now, with \ac{MD} probabilities we further exploit $e^{\theta-\gamma}-1\ge0$ (implied by \e{NJ3}), and $r'(\phi)\le1/\phi-\cot(\phi)$ (implied by  the definition in  \e{GG4}), to write
$$
\eqalign{
b_2(\phi) & \le \frac{(e^{\theta-\gamma}-1) (e^{\theta-r(\phi)}\sinc(\phi)-1)}{(e^{\theta-r(\phi)}-\sinc(\phi))^2} \cr
	& \le \frac{e^{\theta-\gamma}-1}{e^{\theta-r(\phi)}-\sinc(\phi)}\;\sinc(\phi)\cr
	& \le \sinc(\phi)\;e^{r(\phi)-\gamma}\cr
}
\e{DM12}
$$
where the second inequality is a consequence of $\sinc^2(\phi)\le1$, and the third inequality is a consequence of the upper bound in \e{DM8}. For \ac{FA} probabilities an identical upper bound can be identified, but the derivation is different. By considering $e^{\theta-\gamma}-1\le0$ (implied by \e{NJ3}), and $r'(\phi)\ge0$ (implied by the definition in \e{GG4}), we obtain
$$
\eqalign{
b_2(\phi) 	& \le \frac{(1-e^{\theta-\gamma}) (1- e^{\theta-r(\phi)}\cos(\phi))}{(1- e^{\theta-r(\phi)}\cos(\phi))^2} \cr
	& \le \frac{1-e^{\theta-\gamma}}{1- e^{\theta-r(\phi)}\sinc(\phi)} \cr
	& \le \sinc(\phi)\;e^{r(\phi)-\gamma} \cr
}
\e{DM14}
$$
where we assumed $\phi\in[0,\pi)$, and where, for the second inequality, we exploited $\sinc(\phi)\ge\cos(\phi)$, $\phi\in[0,\pi)$. The third inequality in \e{DM14} is instead guaranteed by $1-e^{\theta-\gamma}(1-\sinc^2(\phi))\le \sinc(\phi)\;e^{r(\phi)-\gamma}$, which is a consequence of the lower bound in \e{DM8} and of $\theta\ge-\gamma$ (which, for \ac{FA} probabilities, is in turn guaranteed by the upper bound in \e{NJ3}). Since $b_1(\phi)$ is positive by construction, the above identifies bound
$$
\frac{c(\phi)}{c_0} \le  \overline{b}(\phi) = \frac{\sinc(\fract12\phi)}{1+(r'(\phi))^2} \sqrt{\frac{c_\gamma}{\cosh(r(\phi))}}\;e^{r(\phi)-\gamma}\;.
\e{DM16}
$$
This bound satisfies $ \overline{b}(\phi)\le1$, as we illustrate graphically in \fig{FB33}\Fig[!t]{FB33} for the sake of conciseness. This proves the theorem. Incidentally, we observe that $\overline{b}(\phi)$ behaves as $\sinc(\fract12\phi)$ for small $\gamma\rightarrow0^+$ (rightmost plot in \fig{FB33}), while it behaves as 
$$
\sqrt{\frac{\tan(\fract12\phi)}{\fract12\phi}} \frac1{1+(1/\phi-\cot(\phi))^2}
\e{DM18}
$$
for large $\gamma\rightarrow+\infty$ (leftmost plot in \fig{FB33}).
\end{IEEEproof}}

\begin{IEEEproof}[Proof of Theorem~\ref{leTEO4R}]
For $\snr\rightarrow\infty$ we empirically observe a saturation of the value of $\lambda'$ to a constant depending on block-length and on the value of $P_e$. This correspond, from \e{TE10a}, to the asymptotic equivalence $\gamma \sim \frac12\ln(\snr/\lambda')$. Moreover, we observe that, at the limit, it is $\lambda'>1$. As a consequence, we obtain the asymptotic equivalences
$$
\eqalign{
\theta\sub{MD} & \sim \frac12\ln (\snr\lambda')\cr
\theta\sub{MD} -\gamma & \sim \ln \lambda'\cr
s\sub{MD} & \sim -1\cr
v\sub{MD} &  \sim \lambda' - 1 - \ln (\lambda')\cr
}\quad\eqalign{
\theta\sub{FA} & \sim  \frac12\ln (\lambda'/\snr)\cr
\theta\sub{FA} -\gamma & \sim \ln (\lambda'/\snr)\cr
s\sub{FA} & \sim 1\cr
v\sub{FA} &  \sim \ln (\snr/\lambda') \cr
}
\e{AS2}
$$
We also have (see \e{NM6}) 
$$
\beta(x)=\sqrt{2\alpha(\ln (x))} \sim  -i b(xe^{-\gamma})e^{\gamma/2} \;,
\e{AS4}
$$
with $b(x) = \sqrt{x-1-\ln (x)}$. The asymptotic approximation can be further simplified by limiting the Taylor series of $b(x)$ to the first order, that is $b(x)\sim(x-1)/\sqrt{2}$, to obtain
$$
\beta(x)\sim -i\sqrt{\frac12} (xe^{-\gamma/2}-e^{\gamma/2}) \;,
\e{AS6}
$$
and therefore 
$$
\eqalign{
t(u)&\sim e^\gamma +i\sqrt{2}\,ue^{\gamma/2}\cr
f(u)&\sim-\frac{1}{u -iu_1 }\;,\quad u_1=(1-e^{\theta-\gamma})\sqrt{\fract12e^\gamma}\cr
f_n & \sim - (-iu_1)^{-(n+1)}\cr
c_{2k} & \sim -(-1)^k\frac1{u_1^{2k+1}}\;.
}
\e{AS8}
$$
By substitution in the last of \e{TE4}, and by exploiting \e{TE10c}, we have
$$
g(\gamma)\sim -q\left(u_1\sqrt{n}e^{-\gamma/2}\right)\;,
\e{AS10}
$$
to finally obtain
$$
\eqalign{
g\sub{MD}(\gamma) & \sim  q\Big(\sqrt{n \fract12(\lambda'-1)^2}\Big) \cr
g\sub{FA}(\gamma) & \sim  -q\Big(\sqrt{n/2}\Big)  \;.
}
\e{AS12}
$$
This corresponds to the following asymptotic expression for \e{TE10}
$$
\eqalign{
P\sub{MD} & \sim q\Big(\sqrt{n \fract12(\lambda'-1)^2}\Big)\,e^{ -\frac12n(\lambda'-1-\ln (\lambda'))}\cr
P\sub{FA} & \sim q\Big(\sqrt{n/2}\Big)\,\snr^{-\frac12n} (\lambda')^{\frac12n}\;,
}
\e{AS14}
$$
for some value of the value $\lambda'>1$. The theorem is proved by using \e{AS14} in \e{MO26}, and in \e{NM39}.
\end{IEEEproof}

\bibliographystyle{IEEEtran}
\bibliography{fb}

\end{document}